\documentclass{ws-procs9x6}

\newcommand{\ba}{\begin{eqnarray}}
\newcommand{\ea}{\end{eqnarray}}

\begin{document}

\title{New developments in nuclear supersymmetry: 
pick-up and stripping with susy at ria}

\author{R. BIJKER}

\address{ICN-UNAM, AP 70-543, 04510 Mexico DF, Mexico\\
E-mail: bijker@nucleares.unam.mx}

\maketitle

\abstracts{In this contribution, I discuss the role of symmetries 
and algebraic methods in nuclear structure physics. In particular, 
I review some recent developments in nuclear supersymmetry and 
indicate possible applications for light nuclei in the $sd$- and 
$pf$-shell.}

\section{Introduction}

In recent years, nuclear structure has seen an impressive 
progress in the development of {\em ab initio} methods (no-core 
shell model, Green's Function Monte Carlo, Coupled Clusters, ...), 
mean-field techniques and effective field theories for which the 
ultimate goal is {\em an exact treatment of nuclei utilizing the 
fundamental interactions between nucleons} \cite{bluebook}. 
All involve large scale calculations and therefore   
relie heavily on the available computing power and the development 
of efficient algorithms to obtain the desired results.    

A different approach is that of symmetries and algebraic methods. 
Rather than trying to solve the complex nuclear many-body problem 
numerically, one tries to identify effective degrees of freedom, 
effective or dynamical symmetries, etc. Aside from their esthetic appeal, 
symmetries provide energy formula, selection rules and closed expressions 
for electromagnetic transition rates and transfer strengths which can 
be used as benchmarks to study and interpret the experimental data, 
even if these symmetries may be valid only approximately.   
Symmetries have played an important role in the history of nuclear physics. 
Examples are isospin symmetry, the Wigner supermultiplet theory, 
special solutions to the Bohr Hamiltonian, the Elliott model, 
pseudo-spin symmetries and the dynamical symmetries of the IBM and 
its extensions. 
 
The aim of this contribution is to discuss the role of symmetries and 
algebraic methods in nuclear structure physics. 
In particular, I review some recent results obtained for nuclear 
supersymmetry in the Pt-Au region and discuss some implications 
for light nuclei in the $sd$- and $pf$-shell.  
 
\section{Dynamical supersymmetries in nuclear physics}

Dynamical supersymmetries were introduced \cite{FI} in nuclear physics in 
the context of the Interacting Boson Model (IBM) and its extensions.  
The IBM describes collective excitations in even-even nuclei in 
terms of a system of interacting monopole ($s^{\dagger}$) and quadrupole 
($d^{\dagger}$) bosons, which altogether can be denoted by $b^{\dagger}_{i}$ 
with angular momentum $l=0,2$. The bosons are associated with the number of 
correlated proton and neutron pairs, and hence the number of bosons $N$ is 
half the number of valence nucleons \cite{IBM}. 

In general, the IBM Hamiltonian has to be diagonalized numerically to 
obtain the energy eigenvalues and wave functions. There exist, however, 
special situations in which the eigenvalues can be obtained in closed, 
analytic form. These situations correspond to dynamical symmetries which 
arise, whenever the Hamiltonian is expressed in terms of Casimir invariants 
of a chain of subgroups of $G=U(6)$: the $U(5)$ limit for vibrational nuclei, 
the $SU(3)$ limit for rotational nuclei and the $SO(6)$ limit for 
$\gamma$-unstable nuclei \cite{IBM}. For each one of the dynamical symmetries 
a set of closed analytic expressions has been derived for energies, 
electromagnetic transitions, quadrupole moments and other observables 
of interest which can be used to classify and interpret the available 
experimental data in a qualitative way. 

For odd-mass nuclei the IBM has been extended to include single-particle
degrees of freedom \cite{olaf}. The ensuing Interacting Boson-Fermion Model 
(IBFM) has as its building blocks $N$ bosons with $l=0,2$ and $M=1$ fermion 
($a_j^{\dagger}$) with $j=j_1,j_2,\dots$ \cite{IBFM}. The IBM and IBFM can 
be unified into a supersymmetry (SUSY) \cite{susy} 
\ba
U(6/\Omega) \supset U(6) \otimes U(\Omega) ~,
\ea
where $\Omega=\sum_j (2j+1)$ is the dimension the fermion space. 
In this framework, even-even and odd-even nuclei form the members of a 
supermultiplet which is characterized by $\aleph=N+M$, 
{\em i.e.} the total number of bosons and fermions.  
Supersymmetry distinguishes itself from other symmetries in that it includes, 
in addition to transformations among fermions and among bosons, also 
transformations that change a boson into a fermion and {\em vice versa} 
(see Table~\ref{models}).

\begin{table}[thb]
\centering
\caption[]{\small Algebraic models}
\label{models}
\begin{tabular}{lcccccc}
\hline
Model && Generators && Invariant && Symmetry \\
\hline
IBM &\hspace{0.3cm}& $b^{\dagger}_i b_j$  
&\hspace{0.3cm}& $N$ &\hspace{0.3cm}& $U(6)$ \\
IBFM && $b^{\dagger}_i b_j ~,\; a^{\dagger}_k a_l$ && $N$, $M$ && 
$U(6) \otimes U(\Omega)$ \\
SUSY && $b^{\dagger}_i b_j ~,\; a^{\dagger}_k a_l ~,\; 
b^{\dagger}_i a_k  ~,\; a^{\dagger}_k b_i$ && $\aleph=N+M$ 
&& $U(6/\Omega)$ \\
\hline
\end{tabular}
\end{table}

\section{Supersymmetry in heavy nuclei}

Dynamical nuclear supersymmetries correspond to very special forms 
of the Hamiltonian which may not be applicable to all regions of the 
nuclear chart, but nevertheless many nuclei have been found to provide 
experimental evidence for supersymmetries in nuclei \cite{IBFM,thesis}.  
Especially, the mass region $A \sim 190$ has been a rich source of empirical 
evidence for the existence of (super)symmetries in nuclei. The even-even 
nucleus $^{196}$Pt is the standard example of the $SO(6)$ limit of the 
IBM \cite{so6}. The odd-proton nuclei $^{191,193}$Ir and $^{193,195}$Au 
were suggested as examples of the $Spin(6)$ limit \cite{FI}, in which 
the odd proton is allowed to occupy the $2d_{3/2}$ orbit of the 50-82 
proton shell, whereas the pairs of nuclei $^{190}$Os - $^{191}$Ir, 
$^{192}$Os - $^{193}$Ir, $^{192}$Pt - $^{193}$Au and $^{194}$Pt - $^{195}$Au 
have been analyzed as examples of a $U(6/4)$ supersymmetry \cite{susy}. 

The odd-neutron nucleus 
$^{195}$Pt, together with $^{194}$Pt, were studied in terms of a $U(6/12)$ 
supersymmetry \cite{baha}, in which the odd neutron occupies the $3p_{1/2}$, 
$3p_{3/2}$ and $2f_{5/2}$ orbits of the 82-126 neutron shell. In this case, 
the neutron angular momenta are decoupled into a pseudo-orbital part with 
$\tilde{l}=0,2$ and a pseudo-spin part with $\tilde{s}=\frac{1}{2}$. This  
supersymmetry scheme arises from the equivalence between the values of 
the angular momenta of the pseudo-orbital part and those of the bosons of 
the IBM. 

The concept of nuclear SUSY was extended in 1985 to include the  
neutron-proton degree of freedom \cite{quartet}. In this case, 
a supermultiplet consists of an even-even, an odd-proton, an odd-neutron 
and an odd-odd nucleus. Spectroscopic studies of heavy odd-odd nuclei 
are very difficult due the high density of states. Almost 15 years 
after the prediction of the spectrum of the odd-odd nucleus by nuclear 
supersymmetry, it was shown experimentally that the observed spectrum 
of the nucleus $^{196}$Au is amazingly close to the theoretical one 
\cite{metz}. At present, the best experimental evidence of a supersymmetric 
quartet is provided by the $^{194,195}$Pt and $^{195,196}$Au nuclei as an 
example of the $U(6/12)_{\nu} \otimes U(6/4)_{\pi}$ supersymmetry. 
This supermultiplet is characterized by $\aleph_{\pi}=2$ and $\aleph_{\nu}=5$. 

\begin{figure}[t]
\centering
\includegraphics[height=5cm]{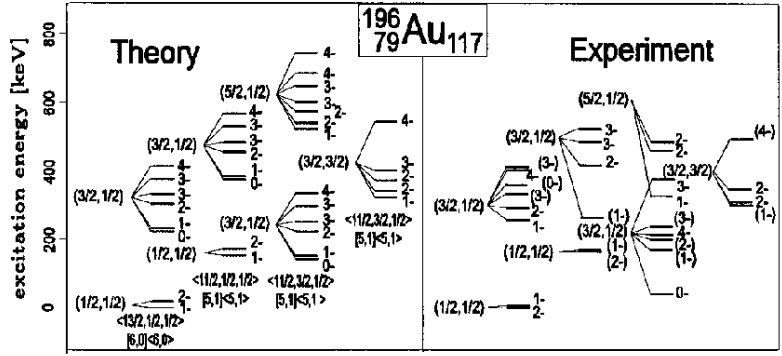}
\caption[]{\small Comparison between the energy spectrum of the negative
parity levels in the odd-odd nucleus $^{196}$Au and that obtained for the
$U(6/12)_{\nu} \otimes U(6/4)_{\pi}$ supersymmetry \cite{au196}.}
\label{odd-odd}
\end{figure}

In this case, the excitation spectra of the supersymmetric quartet of Pt 
and Au nuclei are described simultaneously by the energy formula
\ba
E &=& \alpha \left[ N_1(N_1+5)+N_2(N_2+3)+N_1(N_1+1) \right] 
\nonumber\\
&& + \beta  \left[ \Sigma_1(\Sigma_1+4)+\Sigma_2(\Sigma_2+2)+\Sigma_3^2 \right] 
\nonumber\\
&& + \gamma \left[ \sigma_1(\sigma_1+4)+\sigma_2(\sigma_2+2)+\sigma_3^2 \right] 
\nonumber\\
&& + \delta \left[ \tau_1(\tau_1+3)+\tau_2(\tau_2+1) \right] 
+ \epsilon \, J(J+1) + \eta \, L(L+1) ~.
\label{npsusy}
\ea  
The coefficients $\alpha$, $\beta$, $\gamma$, $\delta$, $\epsilon$ and 
$\eta$ have been determined in a simultaneous fit of the excitation energies 
of the nuclei $^{194,195}$Pt and $^{195,196}$Au \cite{au196}. 
Fig.~\ref{odd-odd} shows the results for the odd-odd nucleus $^{196}$Au. 

\subsection{One-nucleon transfer reactions}

The supersymmetric quartet of nuclei is described by a single Hamiltonian, 
and hence the wave functions, transition and transfer rates are strongly 
correlated. As an example of these correlations, I consider here the case 
of one-proton transfer reactions between the Pt and Au nuclei.    
One-proton transfer reactions between different members of the same 
supermultiplet provide an important test of supersymmetries, since it 
involves the transformation of a boson into a fermion or {\em vice versa}, 
but it conserves the total number of bosons plus fermions. 

The operators that describe one-proton transfer reactions in the 
$U(6/12)_{\nu} \otimes U(6/4)_{\pi}$ supersymmetry are, in lowest order, 
given by 
\begin{eqnarray}
P_{1}^{\dagger} &=& \alpha_1 \left[ -\sqrt{\frac{1}{6}} 
\left( \tilde{s}_{\pi} \times a^{\dagger}_{\pi,3/2} \right)^{(3/2)}  
+\sqrt{\frac{5}{6}} 
\left( \tilde{d}_{\pi} \times a^{\dagger}_{\pi,3/2} \right)^{(3/2)} \right] ~, 
\nonumber\\ 
P_{2}^{\dagger} &=& \alpha_2 \left[ \sqrt{\frac{5}{6}} 
\left( \tilde{s}_{\pi} \times a^{\dagger}_{\pi,3/2} \right)^{(3/2)}  
+\sqrt{\frac{1}{6}} 
\left( \tilde{d}_{\pi} \times a^{\dagger}_{\pi,3/2} \right)^{(3/2)} \right] ~. 
\label{tensor} 
\end{eqnarray}
As a consequence of the selection rules, the operator $P^{\dagger}_{1}$ only 
excites the ground state of the Au nuclei, whereas $P^{\dagger}_{2}$ 
populates, in addition to the ground state, also an excited state 
\cite{JPA}. The ratio of the intensities of one-proton transfer to 
the excited state and to the ground state 
$R=I({\rm gs} \rightarrow {\rm exc})/I({\rm gs} \rightarrow {\rm gs})$ 
does not depend on any parameter, and is given by  
\ba
R_1(^{195}\mbox{Pt} \rightarrow ^{196}\mbox{Au}) 
= R_1(^{194}\mbox{Pt} \rightarrow ^{195}\mbox{Au}) 
&=& 0 ~, 
\nonumber\\
R_2(^{195}\mbox{Pt} \rightarrow ^{196}\mbox{Au}) 
= R_2(^{194}\mbox{Pt} \rightarrow ^{195}\mbox{Au}) 
&=& \frac{9(N+1)(N+5)}{4(N+6)^2} ~, 
\label{ratios} 
\ea
for $P_{1}^{\dagger}$ and $P_{2}^{\dagger}$, respectively. 
The available experimental data from the proton stripping
reactions $^{194}$Pt$(\alpha,t)^{195}$Au and
$^{194}$Pt$(^{3}$He$,d)^{195}$Au \cite{munger} shows that the
$J=3/2$ ground state of $^{195}$Au is excited strongly. 
The relatively small strength to excited $J=3/2$ states suggests 
that the operator $P_{1}^{\dagger}$ of Eq.~(\ref{tensor}) be used 
to describe the experimental data. 

The equality of the ratios of the one-proton transfer reactions 
$^{194}$Pt $\rightarrow$ $^{195}$Au and $^{195}$Pt $\rightarrow$ $^{196}$Au  
is a direct consequence of the supersymmetry classification. 
This prediction has been tested experimentally using the $(^3$He$,d)$ 
reaction on $^{194}$Pt and $^{195}$Pt targets. The results are 
being analyzed at the moment \cite{graw}. 

In addition, in a supersymmetry scheme it is possible to establish 
explicit relations between the intensities of these two transfer 
reactions, {\em i.e.} the one-proton transfer reaction intensities 
between the (ground state of the) Pt and Au nuclei are related by 
\ba
I(^{195}\mbox{Pt} \, \rightarrow \, ^{196}\mbox{Au}) &=& \frac{2L+1}{4} \, 
I(^{194}\mbox{Pt} \, \rightarrow \, ^{195}\mbox{Au}) ~.
\label{corr}
\ea
This correlation can be derived in a general way only using the symmetry 
relations that exist between the wave functions of the even-even, odd-proton, 
odd-neutron and odd-odd nuclei of a supersymmetric quartet. It is important 
to point out, that Eqs.~(\ref{ratios} and (\ref{corr}) are parameter-independent 
predictions which are a direct consequence of nuclear SUSY and which can be 
tested experimentally.  
 
\subsection{Two-nucleon transfer reactions}

Two-nucleon transfer reactions probe the structure of the final nucleus 
through the exploration of two-nucleon correlations that may be present. 
The spectroscopic strengths not only depend on the similarity between 
the states in the initial and final nucleus, but also on the correlation 
of the transferred pair of nucleons. 

In this section, the recent data on the $^{198}$Hg$(\vec{d},\alpha){}^{196}$Au 
reaction \cite{wirth} are compared with the predictions from the 
$U_{\nu}(6/12)\otimes U_{\pi}(6/4)$ supersymmetry. This reaction involves the 
transfer of a proton-neutron pair, and hence measures the neutron-proton 
correlation in the odd-odd nucleus. 
The spectroscopic strengths $G_{LJ}$ 
\begin{equation}
G_{LJ} = | \sum_{j_{\nu} j_{\pi}} g_{j_{\nu}j_{\pi}}^{LJ} 
\left< ^{196}\mbox{Au} \right\| 
( a_{j_{\nu}}^{\dagger} a_{j_{\pi}}^{\dagger} )^{(\lambda)}
\left\| ^{198}\mbox{Hg} \right> |^2 ~, 
\end{equation}
depend on the reaction mechanism via the coefficients $g_{j_{\nu}j_{\pi}}^{LJ}$ 
and on the nuclear structure part via the reduced matrix elements. 

\begin{figure}
\begin{center}
\includegraphics[scale=0.45]{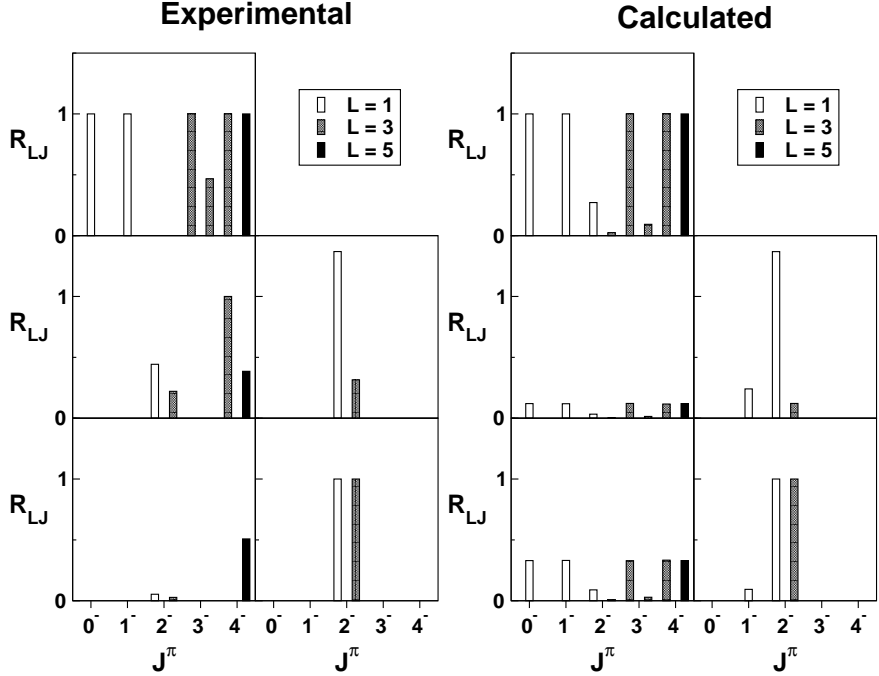}
\end{center}
\caption{Ratios of spectroscopic strengths. 
The two columns in each frame correspond to states with $Spin(5)$ labels 
$(\tau_1,\tau_2)=(\frac{3}{2},\frac{1}{2})$ and 
$(\frac{1}{2},\frac{1}{2})$, respectively. The rows are 
characterized by the labels $[N_1,N_2]$, $(\Sigma_1,\Sigma_2,0)$, 
$(\sigma_1,\sigma_2,\sigma_3)$. From bottom to top we have (i)  
$[6,0]$, $(6,0,0)$, $(\frac{13}{2},\frac{1}{2}, \frac{1}{2})$, (ii) 
$[5,1]$, $(5,1,0)$, $(\frac{11}{2},\frac{1}{2},-\frac{1}{2})$ and (iii) 
$[5,1]$, $(5,1,0)$, $(\frac{11}{2},\frac{3}{2}, \frac{1}{2})$.}
\label{hgau}
\end{figure}

In order to compare with experimental data we calculate the relative 
strengths $R_{LJ}=G_{LJ}/G_{LJ}^{\textrm{ref}}$, 
where $G_{LJ}^{\textrm{ref}}$ is the spectroscopic strength of the
reference state.  Fig.~\ref{hgau} shows the experimental and calculated
ratios $R_{LJ}$. The reference states are easily identified 
since they are normalized to one. 

The ratios of spectroscopic strengths to final states with 
$(\tau_1,\tau_2)=(\frac{3}{2},\frac{1}{2})$  provide a direct test of 
the nuclear wave functions, since they only depend on the nuclear 
structure part \cite{PRL}  
\ba 
R_{2,LJ} &=& \frac{N+4}{15N} ~,
\nonumber\\ 
R_{3,LJ} &=& \frac{2(N+4)(N+6)}{15N(N+3)} ~,
\ea
for different final states corresponding to the left panel of the 
second and third row in Fig.~\ref{hgau}. The numerical values are 
0.12 and 0.33, respectively ($N=5$). 
In general, there is good overall agreement between 
the experimental and theoretical values, especially if we take into
account the simple form of the operator in the calculation of the 
two-nucleon transfer reaction intensities. 

\section{Nuclear supersymmetry in light nuclei}

Dynamical supersymmetries correspond to special solutions of the 
Hamiltonian. They occur whenever the even-even nucleus can be 
described by one of the dynamical symmetries of the IBM, and the 
odd nucleon occupies specific single-particle orbits which lie 
close to the Fermi surface \cite{bijker}. 

In light nuclei, examples may be found in the $sd$-shell for which 
the orbital angular momenta of the $2s_{1/2}$, $1d_{3/2}$ and $1d_{5/2}$ 
orbits match the angular momenta of the bosons \cite{szpikowski}. 
Another area of interest may be the beginning of the $pf$-shell 
where the valence nucleons occupy the orbits $2p_{1/2}$, $1p_{3/2}$ 
and $1f_{5/2}$ which can be treated in a pseudo-spin coupling scheme 
as $\tilde{l}=0,2$ and $\tilde{s}=\frac{1}{2}$. 

\begin{figure}[t]
\centering
\includegraphics[height=9cm]{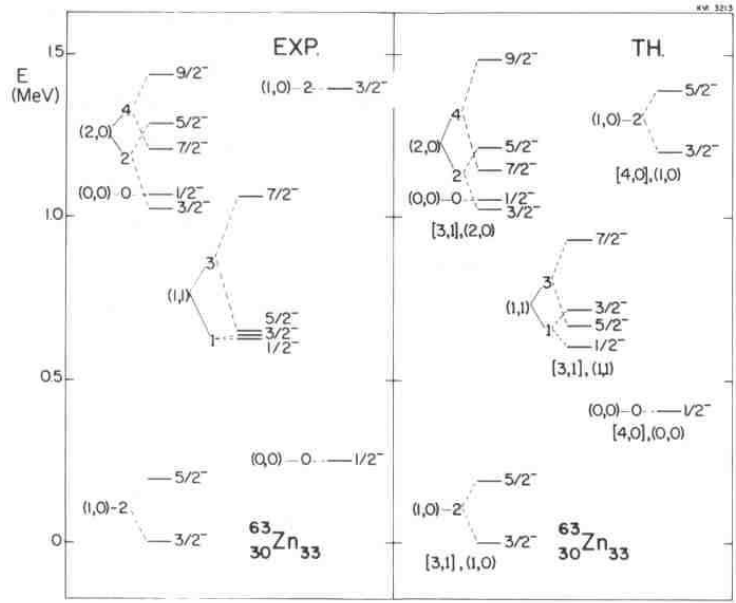}
\caption[]{\small Comparison between the energy spectrum of the negative
parity levels in the nucleus $^{63}$Zn and that obtained for the
$U(36/24)$ supersymmetry \cite{thesis}.}
\label{zinc}
\end{figure}

In heavy nuclei, where the active protons and neutrons occupy different 
major shells, the $s$- and $d$-bosons are associated with 
correlated pairs of identical nucleons with isospin $T=1$ and $M_T=\pm 1$. 
For light nuclei the situation is different, since the valence protons 
and neutrons occupy the same major shell. This observation has led 
to the introduction of an isospin invariant IBM in which the $s$- and 
$d$-bosons can have spin and isospin $(S,T)=(0,1)$ and $(1,0)$ \cite{ibm4}
which leads to the algebraic structure
\ba
U^B(36) \supset U^B_L(6) \otimes SU^B_{ST}(6) \supset U^B_L(6) \otimes SU^B_{ST}(4) ~. 
\ea
The subscripts refer to the orbital part ($L$) and the spin-isospin 
part ($ST$). The group structure of the odd nucleon in the $sd$-shell 
and in the beginning of the $pf$-shell is the same and is given by 
\ba
U^F(24) \supset U^F_L(6) \otimes SU^F_{ST}(4) ~.
\ea
Whereas for the $sd$-shell, $SU_{ST}(4)$ refers to the Wigner supermultiplet 
$SU(4)$ symmetry, for the $pf$-shell it represents the pseudo-$SU(4)$ symmetry 
which follows from the combined invariance in pseudo-spin and isospin \cite{pseudo}. 
It has been shown \cite{pseudo} that the lowlying states of $^{58}$Cu, $^{60}$Zn 
and $^{60}$Ni have good pseudo-$SU(4)$ symmetry. Since the boson and fermion 
chains have the orbital group $U_L(6)$ and the (pseudo)spin-isospin group 
$SU_{ST}(4)$ in common, they may be combined into the supersymmetry
\ba
U(36/24) \supset  U^B(36) \otimes U^F(24) \supset \cdots \supset 
U^{BF}_L(6) \otimes SU^{BF}_{ST}(4) \supset \cdots
\ea
Fig.~\ref{zinc} shows the nucleus $^{63}$Zn as a candidate of a supersymmetry 
in the $pf$-shell \cite{thesis}.
 
\section{Summary, conclusions and outlook}

In this contribution, I have discussed how symmetry methods and algebraic 
models can be used to interpret and help understand the spectroscopic 
properties of atomic nuclei. Nuclear supersymmetry was taken as a 
specific example. Even though most applications have been found in 
medium- and heavy-mass nuclei, there are some interesting possibilities 
for light nuclei as well, especially for the nuclei beyond $^{56}$Ni in 
the beginning of the $pf$-shell where the pseudo-$SU(4)$ symmetry is 
expected to be valid. 

Dynamical supersymmetries provide a set of closed expressions for 
energies, selection rules for electromagnetic transitions and transfer 
reactions which may be used as benchmarks to study and interpret the 
experimental data, even if these symmetries may be valid only in an 
approximate way.   
In such a scheme, a supermultiplet consists of pairs or quartets of nuclei, 
whose properties are described simultaneously by the same form of the 
Hamiltonian, electromagnetic transition operators and transfer operators. 
Therefore, nuclear SUSY predicts explicit correlations between energies 
and electromagnetic transition rates in different nuclei as well as 
between different nucleon transfer reactions which provide a challenge 
and motivation for future experiments. 

The concept of supersymmetry is more general than that of dynamical 
supersymmetry discussed in this contribution. The combination with 
dynamical symmetries has limited the study of nuclear SUSY, 
since dynamical symmetries are rather scarce and only occur in certain 
areas of the nuclear mass table. An example of nuclear SUSY 
without dynamical symmetry is a study of the Ru and Rh isotopes, in 
which an excellent description of the data was obtained by a combination 
of the $U(5)$ and $SO(6)$ dynamical symmetries \cite{once}. 
This opens up the possibility to generalize nuclear SUSY to 
transitional regions of the nuclear mass table, and to extend 
the search for correlations as a result of supersymmetry.  

It goes without saying that symmetry approaches as described in this 
contribution, {\it ab initio} methods, 
mean-field techniques and nuclear effective field theory go hand in hand, 
and provide complementary information about the complex dynamics of the 
nuclear many-body problem.   

\section*{Acknowledgments}

The results for the correlations in the supersymmetric Pt and Au nuclei 
were obtained in collaboration with Jos\'e Barea and Alejandro Frank.   
This work was supported in part by CONACyT.


\begin{thebibliography}{99}

\bibitem{bluebook}
See {\em e.g. RIA Theory Bluebook: A Road Map} (September 2005), 
and various contributions to these proceedings.
 
\bibitem{FI} 
F. Iachello, 
{\em Phys. Rev. Lett.} {\bf 44}, 772 (1980). 

\bibitem{IBM}
F. Iachello and A. Arima, 
{\em The interacting boson model},  
(Cambridge University Press, 1987). 

\bibitem{olaf}
F. Iachello and O. Scholten,
{\em Phys. Rev. Lett.} {\bf 43}, 679 (1979).
 
\bibitem{IBFM}
F. Iachello and P. Van Isacker, 
{\em The interacting boson-fermion model}, 
(Cambridge University Press, 1991). 

\bibitem{susy}
A.B. Balantekin, I. Bars and F. Iachello, 
{\em Phys. Rev. Lett.} {\bf 47}, 19 (1981);\\
A.B. Balantekin, I. Bars and F. Iachello, 
{\em Nucl. Phys.} A {\bf 370}, 284 (1981). 

\bibitem{thesis}
R. Bijker, 
Ph.D. thesis, University of Groningen (1984). 

\bibitem{so6}
J.A. Cizewski, R.F. Casten, G.J. Smith, M.L. Stelts, W.R. Kane, 
H.G. B\"orner and W.F. Davidson,  
{\em Phys. Rev. Lett.} {\bf 40}, 167 (1978);\\
A. Arima and F. Iachello, 
{\em Phys. Rev. Lett.} {\bf 40}, 385 (1978).
 
\bibitem{baha}
A.B. Balantekin, I. Bars, R. Bijker and F. Iachello, 
{\em Phys. Rev.} C {\bf 27}, 1761 (1983).

\bibitem{quartet}
P. Van Isacker, J. Jolie, K. Heyde and A. Frank, 
{\em Phys. Rev. Lett.} {\bf 54}, 653 (1985).

\bibitem{metz}
A. Metz, J. Jolie, G. Graw, R. Hertenberger, J. Gr\"oger, 
C. G\"unther, N. Warr and Y. Eisermann, 
{\em Phys. Rev. Lett.} {\bf 83}, 1542 (1999).

\bibitem{au196}
J. Gr\"oger {\em et al.}, 
{\em Phys. Rev.} C {\bf 62}, 064304 (2000).

\bibitem{JPA}
J. Barea, R. Bijker and A. Frank, 
{\em J. Phys. A: Math. Gen.} {\bf 37}, 10251 (2004).

\bibitem{munger}
M.L. Munger and R.J. Peterson, 
{\em Nucl. Phys.} A {\bf 303}, 199 (1978).

\bibitem{graw}
G. Graw, private communication. 

\bibitem{wirth}
H.-F. Wirth {\em et al.}, 
{\em Phys. Rev.} C {\bf 70}, 014610 (2004).

\bibitem{PRL}
J. Barea, R. Bijker and A. Frank, 
{\em Phys. Rev. Lett.} {\bf 94}, 152501 (2005). 

\bibitem{bijker}
R. Bijker and O. Scholten, 
{\em Phys. Rev.} C {\bf 32}, 591 (1985);\\
R. Bijker and V.K.B. Kota, 
{\em Phys. Rev.} C {\bf 37}, 2149 (1988)

\bibitem{szpikowski}
S. Szpikowski, P. K{\l}osowski and L. Pr\'ochniak, 
{\em Nucl. Phys.} A {\bf 487}, 301 (1988).

\bibitem{ibm4}
J.P. Elliott and J.A. Evans, 
{\em Phys. Lett.} B {\bf 101}, 216 (1981).

\bibitem{pseudo}
P. Van Isacker, O. Juillet and F. Nowacki, 
{\em Phys. Rev. Lett.} {\bf 82}, 2060 (1999).

\bibitem{once}
A. Frank, P. Van Isacker and D.D. Warner, 
{\em Phys. Lett.} B {\bf 197}, 474 (1987).

\end{thebibliography}
\end{document}